\documentclass[prl,twocolumn,letterpaper, superscriptaddress]{revtex4-1}
\usepackage{graphicx}
\usepackage{dcolumn}
\usepackage{bm}
\usepackage{color}
\usepackage{tabularx}
\usepackage{array}
\usepackage{amsmath}
\usepackage{pbox}
\usepackage{stmaryrd}  

\begin{document}

\title{Hybrid magnonics for short-wavelength spin waves facilitated by a magnetic heterostructure}

\author{Jerad Inman}
\affiliation{Department of Physics, Oakland University, Rochester, MI 48309, USA}
\affiliation{Department of Electronic and Computer Engineering, Oakland University, Rochester, MI 48309, USA}

\author{Yuzan Xiong}
\affiliation{Department of Physics, Oakland University, Rochester, MI 48309, USA}
\affiliation{Department of Electronic and Computer Engineering, Oakland University, Rochester, MI 48309, USA}
\affiliation{Materials Science Division, Argonne National Laboratory, Argonne, IL 60439, USA}

\author{Rao Bidthanapally}
\affiliation{Department of Physics, Oakland University, Rochester, MI 48309, USA}

\author{Steven Louis}
\affiliation{Department of Electronic and Computer Engineering, Oakland University, Rochester, MI 48309, USA}

\author{Vasyl Tyberkevych}
\affiliation{Department of Physics, Oakland University, Rochester, MI 48309, USA}

\author{Hongwei Qu}
\affiliation{Department of Electronic and Computer Engineering, Oakland University, Rochester, MI 48309, USA}

\author{Joseph Sklenar}
\affiliation{Department of Physics and Astronomy, Wayne State University, Detroit, MI 48201, USA}

\author{Valentine Novosad}
\affiliation{Materials Science Division, Argonne National Laboratory, Argonne, IL 60439, USA}

\author{Yi Li}
\affiliation{Materials Science Division, Argonne National Laboratory, Argonne, IL 60439, USA}

\author{Xufeng Zhang}
\thanks{xu.zhang@northeastern.edu}
\affiliation{Department of Electrical and Computer Engineering, Northeastern University, Boston, MA 02115, USA}

\author{Wei Zhang}
\thanks{weizhang@oakland.edu}
\affiliation{Department of Physics, Oakland University, Rochester, MI 48309, USA}

\date{\today}

\begin{abstract}

Recent research on hybrid magnonics has been restricted by the long magnon wavelengths of the ferromagnetic resonance modes. We present an experiment on the hybridization of 250-nm-wavelength magnons with microwave photons in a multimode magnonic system consists of a planar cavity and a magnetic bilayer. The coupling between magnon modes in the two magnetic layers, i.e., the uniform mode in Permalloy (Py) and the perpendicular standing spin waves (PSSWs) in YIG,  serves an effective means for exciting short-wavelength PSSWs, which is further hybridized with the photon mode  of  the microwave resonator. The demonstrated magnon-photon coupling approaches the superstrong coupling regime, and can even be achieved near zero bias field.

\end{abstract}

\maketitle

\section{I. Introduction}

Owing to their unique tunability and compatibility, magnons (as elementary excitations of spin waves) have been considered advantageous carriers to bridge different quantum systems \cite{huebl_prl2013,flatte_prl2010,xufeng_sciadv2016,an_prb2020,nakamura_science2015,nakamura_sciadv2017}. This is especially relevant if one uses the yttrium iron garnet, Y$_3$Fe$_5$O$_{12}$ (YIG), due the material's low damping rate and high spin density \cite{serga_jphysd2010}. In particular, \textcolor{black}{in the context of developing hybrid quantum systems,} the \textcolor{black}{coherent} coupling between YIG magnons and microwave photons have exhibited excellent \textcolor{black}{properties such as the ability to} enter the strong coupling regime, opening opportunities for a wide range of useful applications such as nonreciprocal microwave devices \cite{nakamura_apex2019,tqe_2021,jap_2021,hu_prl2019,xufeng_prap2020,harder_prl2018} and dark matter detection \cite{zurek_prl2020,crescini_prl2020,crescini_comphys2020}. 

\begin{figure}[b]
 \centering
 \includegraphics[width=2.8 in]{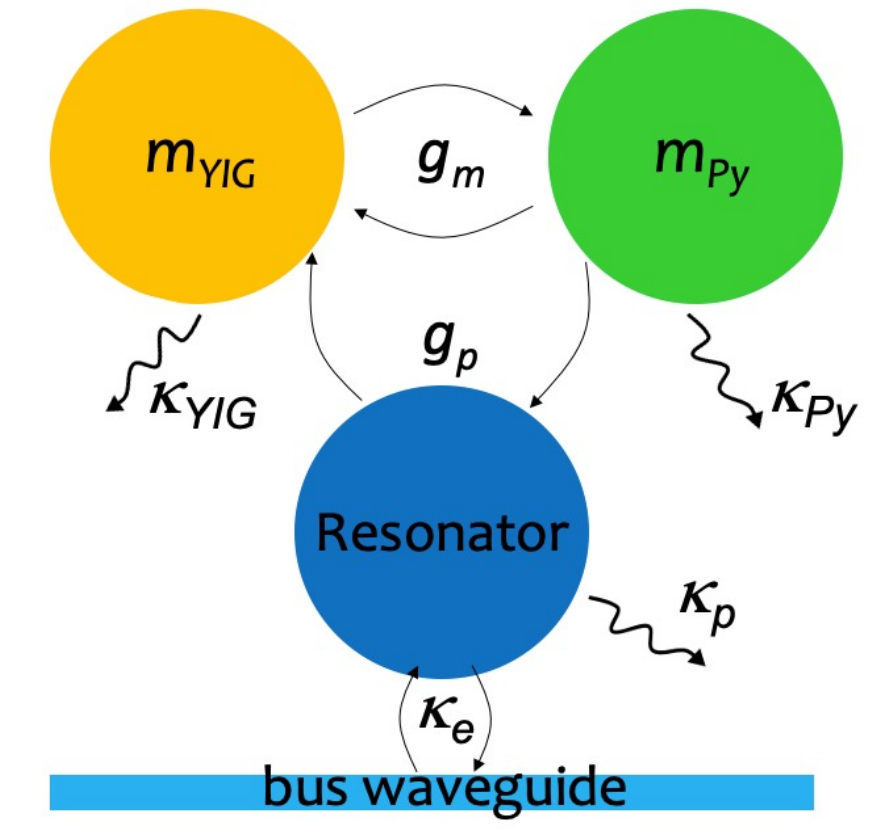}
 \caption{Schematics of the multimode cavity magnonic system consisting of a linearly coupled magnon-magnon system: $m_{Py}$ $\rightleftharpoons$ $m_{YIG}$. $g$, $\kappa_{Py}$, $\kappa_{YIG}$ are the coupling strength and dissipation rates of the respective magnon subsystems. The hybridized magnon modes further couple to a resonator with a dissipation rate $\kappa_{p}$, via a magnon-photon coupling $g_p$. The resonator is coupled to a waveguide bus (feedline), via $\kappa_e$.}
 \label{fig1}
\end{figure}

Nevertheless, \textcolor{black}{most of the works so far on the magnon-microwave photon hybridization has been done in the context of} macroscopic microwave cavity coupled to the uniform magnon mode, i.e., ferromagnetic resonance (FMR) of YIG spheres, \textcolor{black}{with only a few examples on using micro-scale circuit-based resonators and coupling with relatively low-order magnon modes \cite{huebl_prl2013,baity_apl2021,flaig_prb2016,yili_prl2022}}. It is well known that magnons have rich dispersion and can exist in many different mode forms, therefore, it is straightforward to extend such mode hybridization to other types of magnons to take advantage of their unique properties. In particular, when the magnon wavelength is reduced to the nanometer scale, exchange interactions start to set in as the underlying mechanisms for the magnon excitation, which will lead to useful physics and applications. However, the inefficient cavity excitation of spin waves due to their weak coupling to the microwave photons severely limits such  explorations. 

\begin{figure*}[htb]
 \centering
 \includegraphics[width=7.0 in]{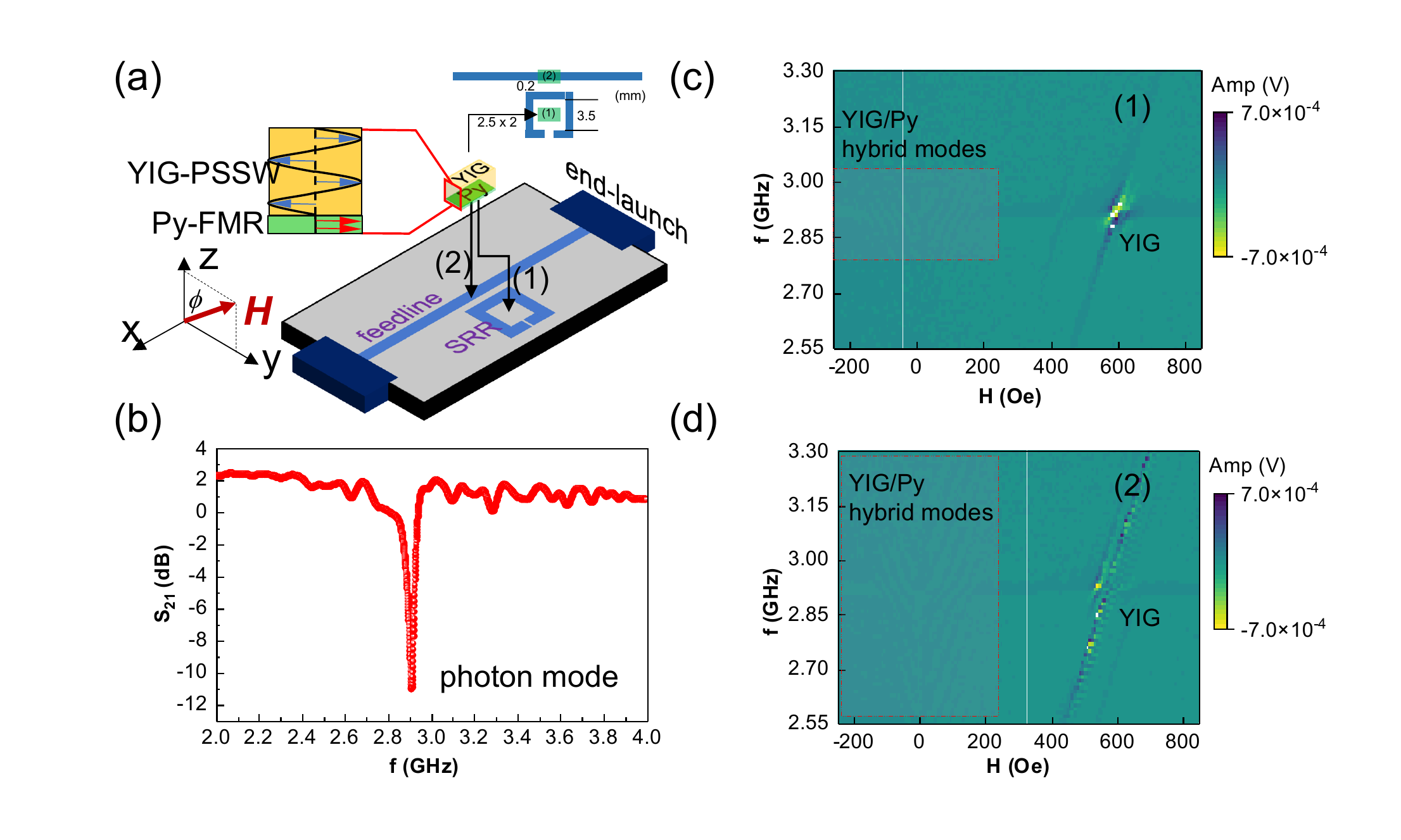}
 \caption{(a) Schematics of the experimental setup, showing the split-ring resonator (SRR) and the YIG/Py bilayer sample. The YIG/Py can be positioned at location (1) and (2), respectively. \textcolor{black}{The dimensions and relative positions between the sample and the SRR are also illustrated.} The magnetic field, $H$ can be applied at an angle $\phi$ with respect to the $z$-axis. (b) The photon mode of the SRR, characterized using a vector-network analyzer. The SRR resonates at a photon frequency, $\omega_p/2\pi=2.912$ GHz and the full-width-half-maximum is $\Delta\omega_p/2\pi=35.2$ MHz. The rough scan of the 2D dispersion contour plot $V$[$f,H$] for the two testing positions in (a): either inside the SRR (c), or on top of the waveguide bus (d). The frequency stepsize is 10 MHz. }
 \label{fig2}
\end{figure*}

Recently, it has been demonstrated that the magnon-magnon coupling in magnetic heterostructures provides an efficient route towards exciting short-wavelength magnons by microwave signals \cite{weiler_prl2018,yu_prl2018,qin_srep2018,yili_magnon2019,xiong_srep2020}. \textcolor{black}{Due to the interfacial exchange torque, high-order magnon modes can be effectively excited by other magnons, whose footprints are then governed by the magnons' dispersion and are not limited by the conventional electromagnetic wave excitations.} Leveraging this approach, here we present an experimental demonstration of a hybrid magnonic system with a magnon wavelength as short as $\sim$ 250 nm. Figure\,\ref{fig1} \textcolor{black}{illustrates} our multimode cavity magnonic system comprised of \textcolor{black}{a bus waveguide coupled to a resonator that is further coupled to a YIG/Py hybrid magnonic system. In our experimental realization, as shown in Fig. \ref{fig2}(a), the resonator is a planar split-ring-resonator (SRR) coupled to a waveguide bus (feedline), and the magnonic system is an exchange-coupled YIG/Py bilayer}. In the Py-biased YIG thin film, the magnons ($m_\mathrm{YIG}^n$) form a series of perpendicular standing spin waves (PSSWs), with wavelengths that can reach the mesoscale regime, at frequencies $\omega_\mathrm{YIG}^n$ with a magnon dissipation rate, $\kappa_\mathrm{YIG}$, where $n$ is the mode order. These modes couple to the uniform mode of Py $m_\mathrm{Py}$ (at frequency $\omega_\mathrm{Py}$ with a magnon dissipation rate $\kappa_\mathrm{Py}$), and the coupling strength is $g_m$. Such hybridized magnon modes further couple to the SRR, with a photon dissipation, $\kappa_p$. The magnon-photon coupling between the hybridized magnon modes and the resonator is $g_p$. Finally, the resonator has a coupling coefficient, $\kappa_e$ to the waveguide bus. The square SRR has an outer dimension of $6.5 \times 6.5$ mm$^2$ and an inner dimension of $3.5 \times 3.5$ mm$^2$. The gap width and distance to the waveguide bus are both 0.2 mm, see Fig.\ref{fig2}(a). The SRR resonates at a photon frequency $\omega_p/2\pi=2.912$ GHz [Fig.\ref{fig2}(b)] and the dissipation rate is $\kappa_p/2\pi=17.6$ MHz, yielding a photon quality(Q)-factor of $\sim$ 83. 

\section{II. Sample and Measurement}

The YIG/Py sample is in the shape of a $2.5 \times 2$ mm$^2$ slab, and consists of a sputtered Py layer (30-nm thick) on top of a 3-$\mu$m-thick, single-sided, commercial single-crystal YIG film grown on double-side-polished Gd$_3$Ga$_5$O$_{12}$ (GGG) substrate using liquid phase epitaxy (LPE). Following our earlier recipes for ensuring a good YIG/Py interfacial exchange coupling \cite{xiong_srep2020}, we used \textit{in-situ} Argon gas rf-bias cleaning for 3 min in the vacuum chamber, to clean the YIG surface before depositing the 30-nm Py layer.

\begin{figure*}[htb]
 \centering
 \includegraphics[width=7.0 in]{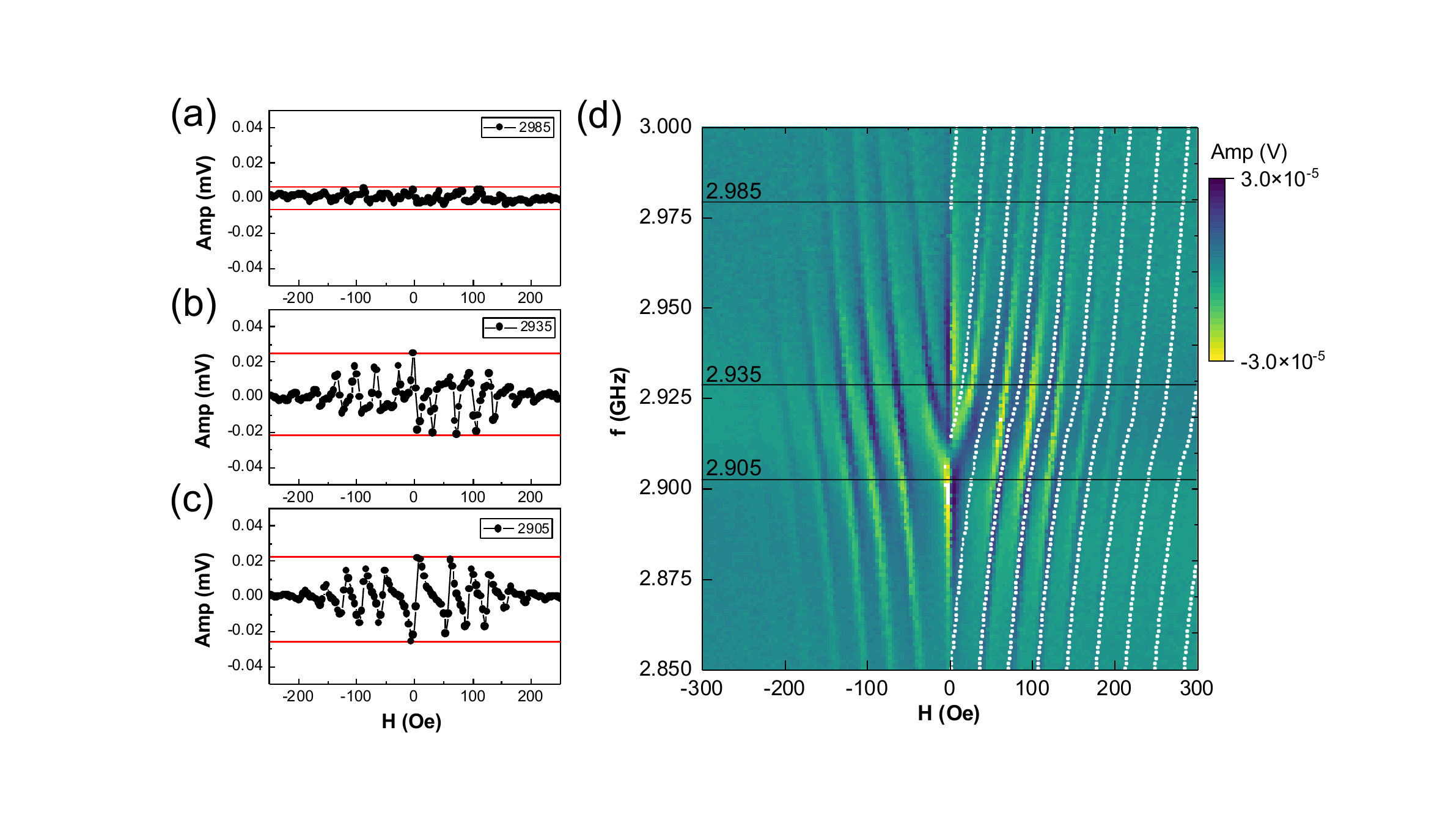}
 \caption{The fine scan datasets at the low field regime (focusing on the hybridized magnon modes) for the testing position (1): inside the SRR. The signal traces at selective frequencies, (a) 2.985, (b) 2.935, and (c) 2.905 GHz, and (d) the corresponding 2D contour plot. The frequency stepsize is 1 MHz. \textcolor{black}{In (d), we also overlay the frequencies of the hybrid modes (dots) extracted from our theoretical calculation by Eq.\ref{Eq:Aout} on the experimental measurement results. } }
 \label{fig3}
\end{figure*}

During the measurement, the samples were \textcolor{black}{mounted in the flip-chip configuration} (Py side facing down) on top of the board for broadband microwave excitation. Two testing configurations were used with the YIG/Py sample placed at two different positions: (1) inside the SRR, and (2) on top of the waveguide bus (feedline, along $x$), as shown in Fig.\ref{fig2}(a). An external bias field, $H$, can be applied at an angle $\phi$ with respect to the normal axis ($z$) using an electromagnet, where $\phi=90^\circ$ corresponds to the field in the plane (along $y$). To achieve the best signal quality, we used the field-modulated FMR technique (as opposed to conventional vector-network analyzer measurement) with a modulation frequency of $\Omega/2\pi=81.57$ Hz (supplied by a lock-in amplifier and provided by a pair of modulation coil) and a modulation field about 1.1 Oe. This modulation level has been tested and chosen not only to be insensitive to the Py resonance profile, but also to give the optimal signal strength of the hybridized magnon modes due to the sharp YIG-PSSW profile. The microwave signal was delivered from a signal generator (10 dBm) to one port of the board. The field-modulated FMR signal was measured from the other port by the lock-in amplifier in the form of a dc voltage, $V$, by using a sensitive rf diode. We swept the bias field, $H$, and at each incremental frequency $f$, to construct the $V$[$f,H$] dispersion contour plots.   

\section{III. Results and Discussions}

We first performed a rough scan ($f$ stepsize = 10 MHz) at larger field range ($\pm 1$ kOe) for the sample placed at the respective (1) and (2) positions with the purpose to identify the nature of the coupling by characterizing the YIG resonance. Figure \ref{fig2}(c) shows the $V$[$f,H$] contour plot for the YIG/Py placed inside the SRR (Position 1) and at a bias field angle $\phi=45^\circ$. This configuration corresponds to the conventional magnon-photon coupling scenario except that we are using a planar resonator instead of a 3D cavity. Accordingly, a standard avoided crossing is observed in the spectra, at $H \sim$ 600 Oe, when the YIG uniform mode crosses the SRR's photon mode. 

Interestingly, a series of additional spin wave modes are also observed, at the lower fields ($\pm 200$ Oe) and even persist down to $H = 0$ Oe, indicating almost a zero-field excitation. These modes, which are much farther away from the YIG's uniform mode, are identified as the hybrid magnon modes caused by the magnon-magnon coupling between the Py uniform mode and the YIG's PSSW modes. In other words, the Py FMR serves as an effective exciter for the PSSWs under its broad resonance envelope \cite{xiong_srep2020}. These modes are only pronounced in a small frequency range, from $\sim$ 2.8 to 3.0 GHz, around the photon resonance. 

We further quantified these hybridized modes by calculating their dispersion relations. Due to the large difference in the magnetization of YIG and Py, the observed PSSW modes that are actively coupled to the Py uniform mode are of a higher indices, $n \sim 20 - 25$, which corresponds to spin-wave wavelengths in the range of $\lambda \sim 240 - 300$ nm. Compared with previous demonstrations which are mostly limited to FMR magnons or low-order magnetostatic magnon modes, our demonstration brings the hybrid magnonic systems into the short wavelength regime. 

On the other hand, Fig.\ref{fig2}(d) shows the $V$[$f,H$] contour plot for the YIG/Py placed at the other position, i.e. on top of the waveguide bus (Position 2) and at the same field angle $\phi=45^\circ$. \textcolor{black}{Similarly, the series of hybrid PSSW modes are observed for bias field strength below 200 Oe. In addition, due to the influence from the traveling wave excitation, such low-field hybrid modes can be detected in a much broader frequency regime as compared to Fig. \ref{fig2}(c). These modes are again attributed to the uniform mode of the Py exciter coupling to the YIG's PSSWs. We also performed such measurements at other magnetic field angles, $\phi=1^\circ, 15^\circ, 30^\circ, 60^\circ, 75^\circ, 90^\circ$, for both the Positions 1 and 2, and the results are included in the Supplemental Materials (SM) \cite{SM}. We can observe such magnon excitation and hybridization at almost all angles. The magnetic field orientation also provides a nice way of tuning the field spacing in the spectrum between different adjacent PSSW modes. } 

To further analyze the hybrid magnon modes and their magnon-photon coupling with the photon mode, we then performed a fine scan ($f$ stepsize = 1 MHz) in a smaller field range ($\pm 400$ Oe) for the sample placed again at the respective (1) and (2) positions. The bias magnetic field is again applied at the field angle $\phi=45^\circ$. Figure \ref{fig3} summarizes the results of the sample located inside the SRR (Position 1). Figure \ref{fig3}(a-c) display example signal traces at three selective frequencies, at: Fig. \ref{fig3}(a), 2.985 GHz, above and away from the photon mode, Fig.\ref{fig3}(b), 2.935 GHz, immediately after the photon mode, and Fig.\ref{fig3}(c), 2.905 GHz, right below the photon mode. At least five hybrid magnon modes can be clearly observed near the photon resonance frequency, see Fig.\ref{fig3}(b-c). The signal amplitude decreases rapidly as the frequency shifts away from the photon mode. For example, the signal at $f=$ 2.985 GHz is almost negligible.

\textcolor{black}{In addition, we note that both the phase and amplitude of the magnon modes are center-symmetric with respect to the field $H = 0$ Oe}, which agrees with the theory of cavity photon-magnon hybridization. Such \textcolor{black}{field-symmetric} magnon dispersion can be also evidenced from the scanned 2D contour plot $V$[$f,H$], see Fig.\ref{fig3}(d). Finally, these hybridized magnon modes show an avoided crossing as their spin wave dispersion passes through the photon mode. In other words, we herein observe the magnon-photon hybridization with a magnon-magnon hybridized mode, and with a significantly reduced magnon wavelength as compared with previous demonstrations.

The hybridization of the PSSW modes in YIG with the FMR mode in Py modifies their intrinsic properties. Previously, these modes cannot be directly readout in the cavity transmission signal because of the weak interaction between the long-wavelength microwave photons and the short-wavelength PSSW magnons. Upon hybridization with the Py FMR mode, the modified PSSW modes have much stronger coupling with the photon mode. On the other hand, the FMR mode of Py is also modified by its interaction with the PSSW modes. However, since \textcolor{black}{(1) the Py FMR mode is very lossy, and (2) our field-modulation strength is only $\sim$ 1 Oe, our lock-in measurement is much less sensitive to the broad Py FMR envelope}. As a result, we can ignore it to simplify our analysis in the following, and we can treat our device as a coupled magnon-photon system, with the magnon mode being the modified PSSW modes in YIG. 

Under this simplified condition, the transmitted RF signal right after the device (before the lock-in detection) can be expressed as \textcolor{black}{\cite{xufeng_prap2021phonon}}:
\begin{equation}
    A_\mathrm{out} = A_0\left[1-\frac{2\kappa_e}{i(\omega_p-\omega)+\kappa_p+\sum_n{\frac{g_n^2}{i(\omega_n-\omega)+\kappa_n}}}\right]\cos{(\omega t)},
\label{Eq:Aout}
\end{equation}
\noindent where $A_0$ in the input amplitude and $\omega$ is the frequency of the microwave tone used in the measurement. It is worth noting that there exist multiple PSSW modes that can interact with the SRR mode, which are accounted for by the summation over the mode order $n$. Here, $\omega_n$, $\kappa_n$, and $g_n$ are the frequency, dissipation rate, and coupling strength (with the SRR mode) of the $n$-th modified PSSW mode, respectively.

\begin{figure}[htb]
 \centering
 \includegraphics[width=3.2 in]{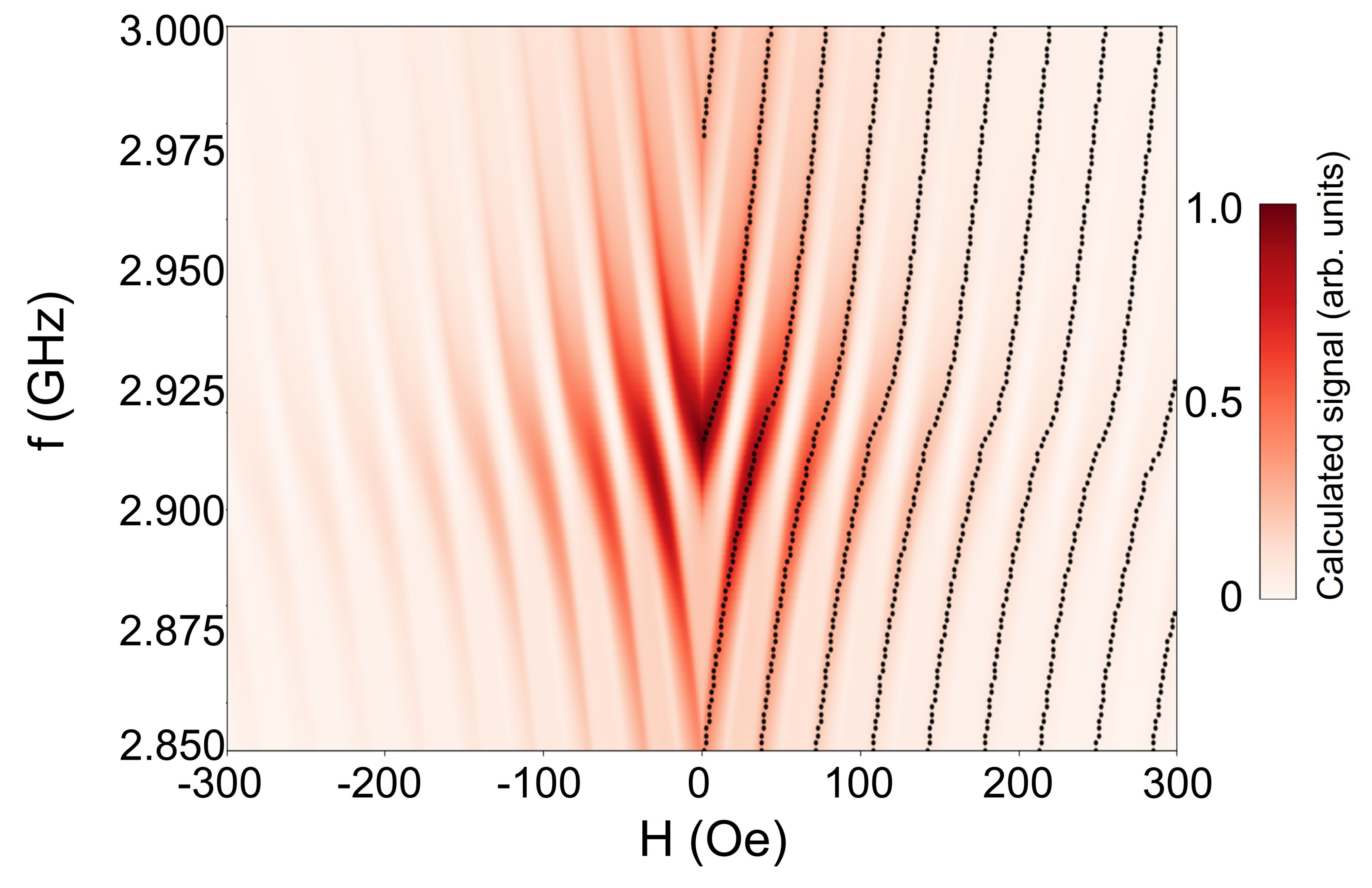}
 \caption{The theoretically-calculated magnon-photon spectra (2D contour plot) in the low field regime using Eq.\,(\ref{Eq:Aout}). \textcolor{black}{The frequencies of the hybrid modes (dots) are extracted and overlaid on the spectrum. } }
 \label{fig4}
\end{figure}

In our measurements, the transmitted RF signal is not directly measured, but instead first sent through a diode envelope detector and then measured by a lock-in amplifier. In our lock-in measurement scheme, the periodically varying bias magnetic field introduces a modulation to the magnon frequency $\omega_n$, which can be simplified as $\omega'_n=\omega_n[1+d\cdot \sin(\Omega t)]$, where the modulation depth $d$ is on the order of $1\%$. The magnonic response of our device is then extracted by detecting the $\Omega$ frequency components in the transmitted RF signals. 

\textcolor{black}{By far-detuning the magnons from the cavity mode, we can determine the frequency and linewidth of each mode by Lorentzian fitting, and the coupling strength is determined through theoretical fitting and calculation.} The lock-in signal spectrum calculated based on parameters extracted from our measurements ($\kappa_p/2\pi= 17.6$ MHz, $\kappa_n/2\pi=  5.5$ MHz, $g_n/2\pi= 22.4$ MHz, PSSW spacing 35.2 Oe) is plotted in Fig.\,\ref{fig4}, which well reproduces the measurement results shown in Fig.\,\ref{fig3}(d). \textcolor{black}{To compare our experiment with the theory, we extracted the frequencies of the hybrid modes from our theoretical calculation and overlaid it on the experimental measurement results in Fig. \ref{fig3}(d), which shows a good agreement with the measured hybrid mode traces. Further,} it is worth noting that the coupling strength $g_n$ is on the same order as the free spectral range (FSR$/2\pi= 112.6$ MHz) of the PSSW modes, indicating that our system is near the super-strong coupling regime, which has been attracting great research interests recently because of its great potential for intriguing physical properties \cite{zhang_jap2016,Kostylev_apl2016}. Finally, we also measured the fine-scan spectra for the sample located on top of the feedline (Position 2), and at some selective field orientations. Similar observations are found and the comprehensive dataset are included in the SM \cite{SM}. 

\section{V. Conclusion}

In summary, we report an experiment of a multimode cavity magnonic  system  comprises  of  a  planar resonator/feedline coupled to a YIG/Py bilayer. The magnon-magnon coupling in  the YIG/Py serves an effective means for exciting short-wavelength PSSW modes. Such short-wavelength modes can further hybridize with the photon mode of the resonator. The demonstrated magnon-photon coupling approaches the super-strong coupling regime, and can even be achieved near zero bias field. \textcolor{black}{Compared with previous demonstrations which are mostly limited to FMR magnons or low-order magnetostatic magnon modes, our demonstration brings the hybrid magnonic systems into the short wavelength regime (towards exchange magnons) and opens opportunities for on-chip integration. In addition, such a spin wave excitation by magnon-magnon coupling to Py is very close to zero bias field, which renders them practically interesting for potentially ``field-free" related applications \cite{yu_2014,liu_2021,oh_2016} or experiments that are sensitive to and/or incompatible with large external magnetic fields such as superconducting circuits. Finally, we note that the same system may also be of interest in studying other topics in the field of hybrid magnonics such as dissipative coupling and level attraction \cite{bai_jmmm2021}, surface spin wave hybridization \cite{jungfleisch_arxiv2022}, nonlinear control of magnon polaritons \cite{hide_arxiv2022}, and coupling-induced transparency \cite{bhoi_arxiv2022}.  }

\section*{Acknowledgment}
The experimental measurement and theoretical modeling at Oakland University was supported by the U.S. National Science Foundation under Grant No. ECCS-1941426 and ECCS-1933301. The sample fabrication and preparation at Argonne National Laboratory was supported by  U.S. DOE, Office of Science, Materials Sciences and Engineering Division. We thank Peng Li, Mahdi Muntasir, and Michael Hamilton for the technical help and useful discussion on the microwave resonator.


\begin{thebibliography}{0}%
\makeatletter
\providecommand \@ifxundefined [1]{%
 \@ifx{#1\undefined}
}%
\providecommand \@ifnum [1]{%
 \ifnum #1\expandafter \@firstoftwo
 \else \expandafter \@secondoftwo
 \fi
}%
\providecommand \@ifx [1]{%
 \ifx #1\expandafter \@firstoftwo
 \else \expandafter \@secondoftwo
 \fi
}%
\providecommand \natexlab [1]{#1}%
\providecommand \enquote  [1]{``#1''}%
\providecommand \bibnamefont  [1]{#1}%
\providecommand \bibfnamefont [1]{#1}%
\providecommand \citenamefont [1]{#1}%
\providecommand \href@noop [0]{\@secondoftwo}%
\providecommand \href [0]{\begingroup \@sanitize@url \@href}%
\providecommand \@href[1]{\@@startlink{#1}\@@href}%
\providecommand \@@href[1]{\endgroup#1\@@endlink}%
\providecommand \@sanitize@url [0]{\catcode `\\12\catcode `\$12\catcode
  `\&12\catcode `\#12\catcode `\^12\catcode `\_12\catcode `\%12\relax}%
\providecommand \@@startlink[1]{}%
\providecommand \@@endlink[0]{}%
\providecommand \url  [0]{\begingroup\@sanitize@url \@url }%
\providecommand \@url [1]{\endgroup\@href {#1}{\urlprefix }}%
\providecommand \urlprefix  [0]{URL }%
\providecommand \Eprint [0]{\href }%
\providecommand \doibase [0]{http://dx.doi.org/}%
\providecommand \selectlanguage [0]{\@gobble}%
\providecommand \bibinfo  [0]{\@secondoftwo}%
\providecommand \bibfield  [0]{\@secondoftwo}%
\providecommand \translation [1]{[#1]}%
\providecommand \BibitemOpen [0]{}%
\providecommand \bibitemStop [0]{}%
\providecommand \bibitemNoStop [0]{.\EOS\space}%
\providecommand \EOS [0]{\spacefactor3000\relax}%
\providecommand \BibitemShut  [1]{\csname bibitem#1\endcsname}%
\let\auto@bib@innerbib\@empty
\end{thebibliography}%


\begin{thebibliography}{19}

\bibitem{huebl_prl2013} H. Huebl, C. W. Zollitsch, J. Lotze, F. Hocke, M. Greifenstein, A. Marx, R. Gross, and S. T. B. Goennenwein, ``High Cooperativity in Coupled Microwave Resonator Ferrimagnetic Insulator Hybrids", Phys. Rev. Lett. \textbf{111}, 127003 (2013). 

\bibitem{flatte_prl2010} O. Soykal and M. E. Flatté, ``Strong field interactions between a nanomagnet and a photonic cavity", Phys. Rev. Lett. \textbf{104}, 077202 (2010).

\bibitem{xufeng_sciadv2016} X.-F. Zhang, C.-L. Zou, L. Jiang, and H. X. Tang, ``Cavity magnomechanics", Sci. Adv. \textbf{2}, e1501286 (2016). 

\bibitem{an_prb2020} K. An, A. N. Litvinenko, R. Kohno, A. A. Fuad, V. V. Naletov, L. Vila, U. Ebels, G. de Loubens, H. Hurdequint, N. Beaulieu, J. Ben Youssef, N. Vukadinovic, G. E. W. Bauer, A. N. Slavin, V. S. Tiberkevich, and O. Klein, ``Coherent long-range transfer of angular momentum between magnon Kittel modes by phonons", Phys. Rev. B \textbf{101}, 060407(R) (2020).

\bibitem{nakamura_science2015} Y. Tabuchi, S. Ishino, A. Noguchi, T. Ishikawa, R. Yamazaki, K. Usami, and Y. Nakamura, ``Coherent coupling between a ferromagnetic magnon and a superconducting qubit, Science \textbf{349}, 405 (2015). 

\bibitem{nakamura_sciadv2017} D. Lachance-Quirion, Y. Tabuchi, S. Ishino, A. Noguchi, T. Ishikawa, R. Yamazaki, and Y. Nakamura, ``Resolving quanta of collective spin excitations in a millimeter-sized ferromagnet", Sci. Adv. \textbf{3}, e1603150 (2017). 

\bibitem{serga_jphysd2010} A. A. Serga, A. V. Chumak, B. Hillebrands, ``YIG magnonics", J. Phys. D.: Appl. Phys. \textbf{43}, 264002 (2010).

\bibitem{nakamura_apex2019} D. Lachance-Quirion, Y. Tabuchi, A. Gloppe, K. Usami, and Y. Nakamura, ``Hybrid quantum systems based on magnonics", Applied Physics Express \textbf{12}, 070101 (2019).

\bibitem{tqe_2021} D. D. Awschalom, C.H.R. Du, R. He, J. Heremans, A. Hoffmann, J. Hou, H. Kurebayashi, Y. Li, L. Liu, V. Novosad, J. Sklenar, S. Sullivan, D. Sun, H. Tang, V. Tyberkevych, C. Trevillian, A. W. Tsen, L. Weiss, W. Zhang, X. Zhang, L. Zhao, and Ch. W. Zollitsch, ``Quantum Engineering With Hybrid Magnonics Systems and Materials", IEEE Trans. Quantum Engineering \textbf{2}, 5500836 (2021). 

\bibitem{jap_2021} Y. Li, W. Zhang, V. Tyberkevych, W.-K. Kwok, A. Hoffmann, V. Novosad, ``Hybrid magnonics: Physics, circuits, and applications for coherent information processing", J. Appl. Phys. \textbf{128}, 130902 (2020).

\bibitem{hu_prl2019} Y. Wang, J. W. Rao, Y. Yang, P.-C. Xu, Y. S. Gui, B. M. Yao, J. Q. You, and C.-M. Hu, ``Nonreciprocity and Unidirectional Invisibility in Cavity Magnonics", Phys. Rev. Lett. \textbf{123}, 127202 (2019).

\bibitem{xufeng_prap2020} X. Zhang, A. Galda, X. Han, D. Jin, and V. M. Vinokur, ``Broadband Nonreciprocity Enabled by Strong Coupling of Magnons and Microwave Photons", Phys. Rev. Applied \textbf{13}, 044039 (2020). 

\bibitem{harder_prl2018} M. Harder, Y. Yang, B.M. Yao, C.H. Yu, J.W. Rao, Y.S. Gui, R.L. Stamps, and C.-M. Hu, ``Level Attraction Due to Dissipative Magnon-Photon Coupling", Phys. Rev. Lett. \textbf{121}, 137203 (2018). 

\bibitem{zurek_prl2020} T. Trickle, Z. Zhang, and K. M. Zurek, ``Detecting Light Dark Matter with Magnons", Phys. Rev. Lett. \textbf{124}, 201801 (2020). 

\bibitem{crescini_prl2020} N. Crescini, D. Alesini, C. Braggio, G. Carugno, D. D’Agostino, D. Di Gioacchino, P. Falferi, U. Gambardella, C. Gatti, G. Iannone, C. Ligi, A. Lombardi, A. Ortolan, R. Pengo, G. Ruoso, and L. Taffarello (QUAX Collaboration), ``Axion Search with a Quantum-Limited Ferromagnetic Haloscope", Phys. Rev. Lett. \textbf{124}, 171801 (2020).

\bibitem{crescini_comphys2020} N. Crescini, C. Braggio, G. Carugno, R. Di Vora, A. Ortolan, and G. Ruoso, ``Magnon-driven dynamics of a hybrid system excited with ultrafast optical pulses", Communications Physics \textbf{3}, 164 (2020). 

\bibitem{baity_apl2021} P. G. Baity, D. A. Bozhko, R. Macedo, W. Smith, R. C. Holland, S. Danilin, V. Seferai, J. Barbosa, R. R. Peroor, S. Goldman, U. Nasti, J. Paul, R. H. Hadfield, S. McVitie, and M. Weides, ``Strong magnon-photon coupling with chip-integrated YIG in the zero-temperature limit", Appl. Phys. Lett. \textbf{119}, 033502 (2021). 

\bibitem{flaig_prb2016} H. Maier-Flaig, M. Harder, R. Gross, H. Huebl, and S. T. B. Goennenwein, ``Spin pumping in strongly coupled magnon-photon systems", Phys. Rev. B \textbf{94}, 054433 (2016). 

\bibitem{yili_prl2022} Y. Li, V. G. Yefremenko, M. Lisovenko, C. Trevillian, T. Polakovic, T. W. Cecil, P. S. Barry, J. Pearson, R. Divan, V. Tyberkevych, C. L. Chang, U. Welp, W.-K. Kwok, and V. Novosad, ``Coherent Coupling of Two Remote Magnonic Resonators Mediated by Superconducting Circuits", Phys. Rev. Lett. \textbf{128}, 047701 (2022). 


\bibitem{weiler_prl2018} S. Klingler, V. Amin, S. Gepr¨ags, K. Ganzhorn, H. Maier-Flaig, M. Althammer, H. Huebl, R. Gross, R. D. McMichael, M. D. Stiles, S. T. B. Goennenwein, and M. Weiler, ``Spin-Torque Excitation of Perpendicular Standing Spin Waves in Coupled YIG/Co Heterostructures", Phys. Rev. Lett. \textbf{120}, 127201 (2018).

\bibitem{yu_prl2018} J. Chen, C. Liu, T. Liu, Y. Xiao, K. Xia, G. E. W. Bauer, M. Wu, and H. Yu, ``Strong Interlayer Magnon-Magnon Coupling in Magnetic Metal-Insulator Hybrid Nanostructures", Phys. Rev. Lett. \textbf{120}, 217202 (2018).

\bibitem{qin_srep2018} H. Qin, S. J. Hamalainen, and S. van Dijken, ``Exchange-torque-induced excitation of perpendicular standing spin waves in nanometer-thick YIG films", Sci. Rep. \textbf{8}, 5755 (2018).

\bibitem{yili_magnon2019} Y. Li, W. Cao, V. P. Amin, Z. Zhang, J. Gibbons, J. Sklenar, J. Pearson, P. M. Haney, M. D. Stiles, W. E. Bailey, V. Novosad, A. Hoffmann, and W. Zhang, ``Coherent spin pumping in a strongly coupled magnon-magnon hybrid system", Phys. Rev. Lett. \textbf{124}, 117202 (2020).

\bibitem{xiong_srep2020} Y. Xiong, Y. Li, M. Hammami, R. Bidthanapally, J. Sklenar, X. Zhang, H. Qu, G. Srinivasan, J. Pearson, A. Hoffmann, V. Novosad, W. Zhang, "Probing magnon–magnon coupling in exchange coupled Y3Fe5O12/Permalloy bilayers with magneto-optical effects", NPG Sci. Rep. \textbf{10}, 12548 (2020). 

\bibitem{SM} See Supplemental Material at http://link.aps.org/supplemental/XXXXX for additional supplemental datasets measured at different field angles. 

\bibitem{xufeng_prap2021phonon} J. Xu, C. Zhong, X. Zhou, X. Han, D. Jin, S. K. Gray, L. Jiang, and X. Zhang, ``Coherent Pulse Echo in Hybrid Magnonics with Multimode Phonons", Phys. Rev. Applied \textbf{16}, 024009 (2021).

\bibitem{zhang_jap2016} X. Zhang, C. Zou, L. Jiang, and H. X. Tang, ``Superstrong coupling of thin film magnetostatic waves with microwave cavity", J. Appl. Phys. \textbf{119}, 023905 (2016).

\bibitem{Kostylev_apl2016} N. Kostylev, M. Goryachev, and M. E. Tobar, ``Superstrong coupling of a microwave cavity to yttrium iron garnet magnons", Appl. Phys. Lett. \textbf{108}, 062402 (2016).

\bibitem{yu_2014} G. Yu, P. Upadhyaya, Y. Fan, J. G. Alzate, W. Jiang, K. L. Wong, S. Takei, S. A. Bender, L.-T. Chang, Y. Jiang, M. Lang, J. Tang, Y. Wang, Y. Tserkovnyak, P. K. Amiri, and K. L. Wang, ``Switching of perpendicular magnetization by spin–orbit torques in the absence of external magnetic fields", Nature Nanotechnology \textbf{9}, 548 (2014).

\bibitem{liu_2021} L. Liu, C. Zhou, X. Shu, C. Li, T. Zhao, W. Lin, J. Deng, Q. Xie, S. Chen, J. Zhou, R. Guo, H. Wang, J. Yu, S. Shi, P. Yang, S. Pennycook, A. Manchon, and J. Chen, ``Symmetry-dependent field-free switching of perpendicular magnetization", Nature Nanotechnology \textbf{16}, 277 (2021). 

\bibitem{oh_2016} Y. W. Oh, S. C. Baek, Y. M. Kim, H. Y. Lee, K.-D. Lee, C.-G. Yang, E.-S. Park, K.-S. Lee, K.-W. Kim, G. Go, J.-R. Jeong, B.-C. Min, H.-W. Lee, K.-J. Lee, and B.-G. Park, ``Field-free switching of perpendicular magnetization through spin–orbit torque in antiferromagnet/ferromagnet/oxide structures", Nature Nanotechnology \textbf{11}, 878 (2016). 

\bibitem{bai_jmmm2021} Z. Lu, X. Mi, Q. Zhang, Y. Sun, J. Guo, Y. Tian, S. Yan, and L. Bai, ``Interference induced microwave transmission in the YIG-microstrip cavity system", J. Magn. Magn. Mater. \textbf{540}, 168457 (2021).

\bibitem{jungfleisch_arxiv2022} M. T. Kaffash, D. Wagle, A. Rai, T. Meyer, J. Q. Xiao, and M. B. Jungfleisch, ``Direct probing of strong magnon-photon coupling in a planar geometry", arXiv:2202.12696 (2022).

\bibitem{hide_arxiv2022} O. Lee, K. Yamamoto, M. Umeda, Ch. W. Zollitsch, M. Elyasi, T. Kikkawa, E. Saitoh, G. E. W. Bauer, and H. Kurebayashi, ``Nonlinear control of magnon polaritons", arXiv:2201.10889 (2022).

\bibitem{bhoi_arxiv2022} B. Bhoi, B. Kim, H. Jeon, and S. Kim, ``Coexistence of coupling-induced transparency and absorption of transmission signals in magnon-mediated photon-photon coupling", arXiv:2202.02667 (2022).

\end{thebibliography}
\end{document}